\documentclass[letterpaper]{article}
\usepackage{aaai}
\usepackage{times}
\usepackage{helvet}
\usepackage{courier}
\usepackage{booktabs}
\usepackage{caption}
\usepackage{subfig}
\usepackage{graphicx}
\usepackage{comment}
\begin{document}
%

\title{Friends, Strangers, and the Value of Ego Networks for Recommendation}
\author{
Amit Sharma\\
Dept. of Computer Science\\
Cornell University\\
Ithaca, NY 14853 USA\\
asharma@cs.cornell.edu
\And
Mevlana Gemici\\
       Dept. of Computer Science\\
       Cornell University\\
       Ithaca, NY 14853 USA\\
       mcg74@cornell.edu
\And
Dan Cosley\\
       Information Science\\
       Cornell University\\
       Ithaca, NY 14853 USA\\
       danco@cs.cornell.edu
}

\maketitle

\begin{abstract}
\begin{quote}
Two main approaches to using social network information in recommendation have emerged: augmenting collaborative filtering with social data and algorithms that use only ego-centric data. We compare the two approaches using movie and music data from Facebook, and hashtag data from Twitter. We find that recommendation algorithms based only on friends perform no worse than those based on the full network, even though they require much less data and computational resources. 
Further, our evidence suggests that locality of preference, or the non-random distribution of item preferences in a social network, is a driving force behind the value of incorporating social network information into recommender algorithms.  When locality is high, as in Twitter data, simple \textit{k}-nn recommenders do better based only on friends than they do if they draw from the entire network.
These results help us understand when, and why, social network information is likely to support recommendation systems, and show that systems that see ego-centric slices of a complete network (such as websites that use Facebook logins) or have computational limitations (such as mobile devices) may profitably use ego-centric recommendation algorithms. 
\end{quote}
\end{abstract}

\noindent An increasingly common approach in recommender systems research has been to use network ties in recommendation.  One main strategy for this is to augment collaborative filtering algorithms with social data \cite{ma11,konstas09}.
However, in many cases, computation on the full dataset may be undesirable or impossible. For instance, most online social networks reveal only a user's first degree connections through their APIs, meaning that many third-party recommendation providers only have access to the immediate friends of a user. 

Thus, another main strategy for using social information has been to focus on making recommendations using only ego-centric slices of the network \cite{jilinchen10,guy09}.  Ego-centric algorithms have clear advantages in terms of the data and computational resources required; however, not much is known about the relative performance of the two approaches. 

In this paper, we empirically compare these two approaches to social recommendation using datasets from Facebook and Twitter. 
We find that algorithms using only friends' data are comparable in performance to those using the full network---and for Twitter data, they perform better.

We hypothesize that this surprising result may be because of the prevalence of \textit{locality} in these networks. By this we mean that the preferences of people close to each other in the network are more similar than one would expect if they were independently distributed, an intuition called out by Ma et al. in early social recommendation work \cite{ma08}.  To explore this, we develop a set of metrics that characterize the extent of locality in a network.  For our datasets, we find that Twitter data exhibits the highest locality with all metrics; that, combined with the fact that ego-centric recommendations are most effective on Twitter, suggests that preference locality is an important factor in the success of such recommendations. 

Overall, our results suggest ego-centric approaches may be good enough for recommendation in social contexts, to the extent that locality exists. 
Further, the extent of locality varies between datasets, which suggests that the payoff of using social data may also vary with domain.  For practitioners, our proposed locality metrics can be helpful to characterize the expected benefits of social recommendation (at least in terms of recommender performance).  For researchers, our results point to the need to better understand the distribution of preferences over individuals and networks in developing effective recommendation algorithms.

\paragraph{Data sources.} 
We use three datasets: two from Facebook (one for movies, one for music) and one from Twitter (hashtag use). The Facebook datasets were collected as a part of a user study involving university students \cite{sharma13}, while the Twitter data was collected using their public API \cite{mcauley12}.

The datasets have an ego-centric structure: each \textit{core user} is guaranteed to have all his first-degree connections. 
The preference data is a set of user-item pairs, where items are movies or musical artists Liked in Facebook or hashtags used in Twitter. We use the term \textit{friends} to refer to first-degree connections or followees of a user,
\textit{ego networks} to refer to a core user and her friends, \textit{non-friends} to refer to everyone not in a given ego network, \textit{items} to refer to movies, artists, or hashtags, and \textit{likes} to refer to Likes on Facebook or usage of a hashtag.

\begin{figure}[t]
        \centering
        \includegraphics[width=0.45\textwidth]{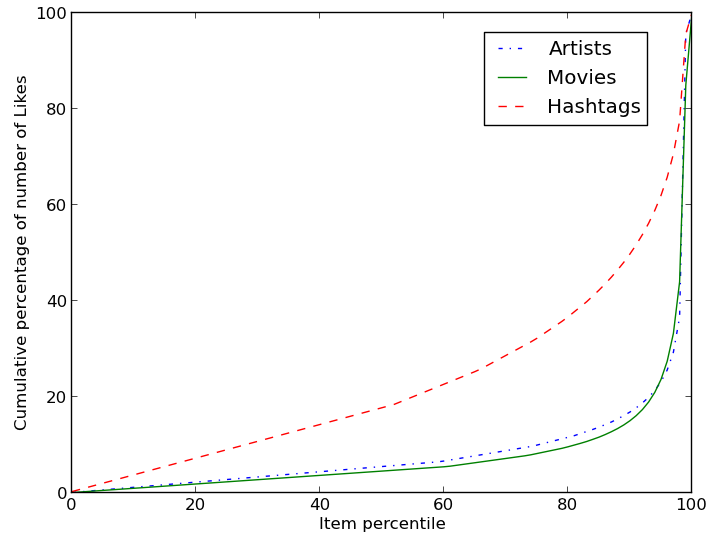}
        
        \caption{Distribution of item likes for the three domains, showing the cumulative percentage of total likes for items in the $k$th percentile of popularity. Compared to hashtags, likes for artists and movies are skewed toward popular items.}
        \label{fig:distr_items}
\end{figure}

\begin{table}
	\centering
	\begin{tabular}{lrrr}
		\toprule[0.15em]
		\textbf{} & \textbf{Artists} & \textbf{Movies} & \textbf{Hashtags} \\ \midrule
		Total users 			& 63230 & 51365 & 69414\\
		Total core users		& 153 & 149 & 935\\
		Friends/user ($\mu;\sigma$)	& (499; 341) & (352; 156) & (126; 61) \\
		Total items				& 139986 & 78244 & 214941 \\
		Total likes				& 1289340 & 873261 & 1230169\\
		Likes/user ($\mu;\sigma$)	& (20; 34) & (17; 33) & (17; 17) \\ 
		Likes/item ($\mu;\sigma$)	& (9; 117) & (11; 96) & (6; 29) \\ 
		\bottomrule[0.15em]
		
	\end{tabular}
	\caption{Overview of datasets. Both the Facebook artists and movies datasets have a higher friend average and average number of likes per item compared to Twitter.}
	\label{tab:prelim_stats}
\end{table}
Table~\ref{tab:prelim_stats} shows the statistics for the three datasets. The two Facebook networks 
have both a higher average number of friends than Twitter and a higher average likes per item. The distribution of likes for items (Figure~\ref{fig:distr_items}) shows that likes for artists and movies are concentrated toward the most popular items: the 10\% most popular artists and movies receive around 5/6 of the likes, versus about half for hashtags.

Artists and movies have fairly similar profiles with a long tail and an uneven distribution of popularity.  We suspect that this happens because of exogenous effects such as media exposure for the most popular artists and movies that generates large amounts of attention for a relatively small number of items.  Hashtag use in Twitter, on the other hand, is much more evenly distributed, perhaps because hashtags receive less exposure outside of Twitter itself and so must spread largely through the Twitter network.

\section{Goodness of ego network recommendations}	
We now turn to our main question, around how ego network-based recommendations compare to recommendations that use the full dataset.
We use a simple \textit{k}-nn recommender as a reasonable baseline algorithm that also allows us to directly discuss similarity between users\footnote{We tried a number of algorithms; the patterns were similar to \textit{k}-nn, so, like Fermat, we omit the details due to lack of space.}.

\paragraph{\textit{k}-nn similarity.} 
Recommender algorithms typically choose the most similar neighbors, so we first look at how a user's $k$ most similar friends compare to the most similar $k$ non-friends in the network.  We use the Jaccard similarity coefficient,  a common measure for unary ratings in recommender systems \cite{candillier08}. For two users $u_1$ and $u_2$, it is given by:
$$JS = \frac{|Likes(u_1) \cap Likes(u_2)|}{|Likes(u_1) \cup Likes(u_2)|}$$  

\begin{figure}[t]
        \centering
        \includegraphics[scale=0.45]{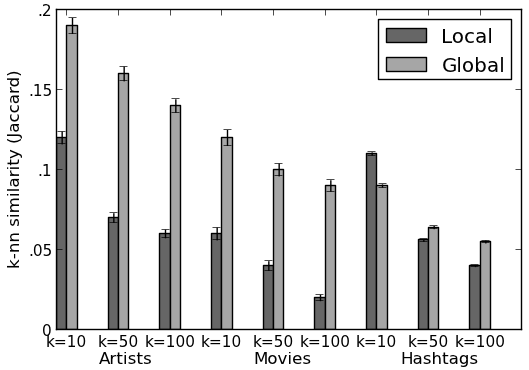}                          
        \caption{Average Jaccard similarity between a user, her top-\textit{k} friends, and her top-\textit{k} non-friends. For the Facebook datasets, non-friends are more similar at all values of $k$; friends appear to be more informative for hashtags.}
        \label{fig:knn_sim}
\end{figure}

Figure~\ref{fig:knn_sim} shows the results. As expected, average similarity decreases as \textit{k} increases. For artists and movies, the top-\textit{k} friends are less similar than the top-\textit{k} non-friends.  Hashtags offer a different scenario: 
 a core user's top 10 friends are more similar than the top 10 others.

\paragraph{Recommendation quality.} We next evaluate how useful friends are compared to the whole network for recommendation.  We divide each core user's preference data into 70:30 train-test splits (using 30\% test items means even users with few likes have at least one in the test set) and make top-10 recommendation lists using items liked by the $k$ nearest neighbors, weighted by Jaccard similarity. For evaluation, we use normalized discounted cumulative gain (NDCG), a common metric used to compare ranked results:
$$\textsc{NDCG} = \frac{Rel_1 + \sum_{2..N} Rel_i / \log_{2}{i}}{1 + \sum_{2..N} 1 / \log_2{i}}$$
where $N=\{min{(10, |TestSet|)}\}$ and $Rel_i=1$ if the $i$th ranked item is in the test set and $0$ otherwise.

\begin{figure}[t]
        \centering
       \includegraphics[scale=0.45]{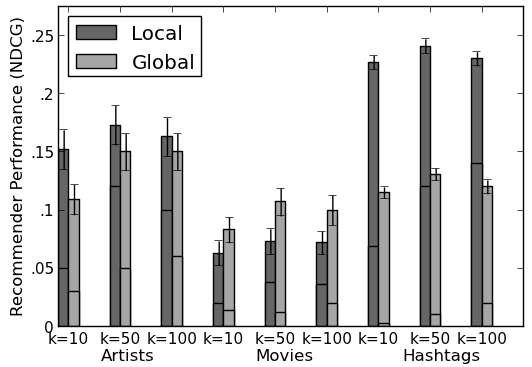}

        \caption{NDCG for \textit{k}-nn recommender using only friends versus only non-friends.  Friends are better for artists and hashtags, worse for movies.  The black marker within each bar represents a recommender choosing people randomly. Error bars represent standard deviation of the mean.}
        \label{fig:knn_recs}
\end{figure}

To focus on the effect of friends, we first compare ego-centric recommendations based on only friends to recommendations that use only non-friends from the full network.  Figure~\ref{fig:knn_recs} shows the results averaged across 10 random 70:30 splits.  Compared to non-friends, friends are better for artists, worse for movies, and much better for hashtags.  For both artists and movies, performance differences between friends and non-friends are much lower than might be expected based on the large differences in \textit{k}-nn similarities from Figure~\ref{fig:knn_sim}.

In practice, algorithms would include friends as part of the full network as well.  Using the full network (including friends) improves performance (Table~\ref{tab:knn_recs2}), but ego-centric recommendations are still comparable, and for hashtags, the ego network-based recommender performs slightly better than the one using the full network.

\begin{table}[t]
	\centering
	\begin{tabular}{llll}
		\toprule[0.15em]
		\textbf{} & \textbf{Friends} & \textbf{Non-Friends} & \textbf{Full Network}\\ 
		\midrule
		Artists 	& 0.17 & 0.15 & 0.17\\
		Movies		& 0.07 & 0.11 & 0.11\\
		Hashtags	& 0.24 & 0.13 & 0.22\\
		\bottomrule[0.15em]
		
	\end{tabular}
	\caption{NDCG values for 50-nn recommenders using the ego network, non-friends, and the full network.  The full network recommender shows improved performance, but is still comparable to the algorithm based only on ego network data.}

	\label{tab:knn_recs2}
\end{table}

\section{Locality of preferences}
We now look at properties of the datasets that might explain these results, focusing on comparing ego networks with the full dataset in terms of similarity, sparsity, and item coverage.  In all three cases, we find evidence that preferences are in fact unevenly distributed, as proposed by \citeauthor{ma08} \shortcite{ma08}.

\begin{table}[b]
	\centering
	\begin{tabular}{lllll}
		\toprule[0.15em]
		& \multicolumn{2}{c}{\textbf{Similarity}} & \multicolumn{2}{c}{\textbf{Sparsity}} \\
		\cmidrule{2-3}
		\cmidrule{4-5}
		& \multicolumn{1}{c}{Friends} & \multicolumn{1}{c}{Non-Friends} & \multicolumn{1}{c}{Ego} & \multicolumn{1}{c}{Network}\\ 
		\midrule
		Artists 	 & 0.040 & 0.023 & 0.73\% & 0.01\%\\
		Movies	 & 0.020 & 0.013 & 0.87\% & 0.02\% \\
		Hashtags  & 0.043 & 0.002 & 3.60\% & 0.01\%\\
		\bottomrule[0.15em]
	\end{tabular}
	\caption{Matrix-based locality metrics.  ``Similarity'' is the average Jaccard similarity between a user and her $k$ friends or $k$ randomly selected non-friends. ``Sparsity'' refers to ratings density in the user-item matrix in ego networks versus the network as a whole.}
	\label{tab:sparsity}
\end{table}

\paragraph{Similarity.} We first look at how similar a person is to his friends versus randomly selected people in the full dataset.
We measure this directly by comparing the average Jaccard similarity between a core user and his friends versus that user and an equal number of randomly chosen non-friends.
Table~\ref{tab:sparsity} shows the results averaged over 10 random sets of non-friends. In all three datasets, friends are in fact more similar.  The effect is strongest with hashtags, consistent with our expectation of relatively strong endogenous effects in the adoption of hashtags compared to artists or movies.  These results suggest that for any set of people chosen at random from a user's friends or the full network, friends are expected to be more similar.

\paragraph{Sparsity.} Ratings sparsity is common in preference datasets, including ours, which have an average user-item matrix density of 0.02\% or less.  However, preference locality suggests that this sparsity should be unevenly distributed.  Table~\ref{tab:sparsity} shows that this is in fact the case: ratings for the items in a given ego network are two orders of magnitude denser than in the full dataset. Again, the effect is stronger for hashtags than for movies or artists, reflecting their higher locality. 

\paragraph{Ego Coverage Metrics.} Another way to think about preference locality is to look at coverage, or what percentage of items can be recommended to a user.
Intuitively, as an item's preferences are more localized, it will be liked in fewer ego networks, reducing coverage.  There are a number of ways we might formalize this notion. 

The simplest approach is to look at the percentage of ego networks in which at least one person likes a given item.  Averaging this over all items gives us a measure of ego network coverage: what percent of possible ego network-item pairs exist in the dataset?  We then subtract from 100 so that higher numbers correspond to increased locality, and call this the \textit{Uncovered Ego} of the network.

\textit{Uncovered Ego}, however, does not account for item popularity.  For instance, items with one like will look maximally local---true but uninformative---while an item with many likes should appear in many networks, and thus appear less local even if its preferences are more concentrated than we expect.  Thus, we might want to account for the expected number of ego networks an item should appear in.  To do this, we create a network with identical friend connections and number of likes per node, but with the items randomly distributed (subject to the constraint that each node can only like a given item once). Dividing the number of ego networks that contain a given item in the randomized network by the number of ego networks that contain it the real network gives a measure of how much preferences deviate from what we would expect if they were distributed without reference to the friend network.  We call this metric \textit{Random Item/Ego}, with higher numbers indicating greater locality.  

In some ways, that approach is too random because it doesn't account for patterns of individual preferences---that, as Amazon reminds us, people who bought X also tend to buy Y.  One way to account for this is to randomize at the network level rather than the item level: keeping the same number of friends for each core user and the same itemsets, but randomly reassigning the friend links.  We call this metric \textit{Random Friend/Ego}.

\begin{table}[tb]
	\centering
	\begin{tabular}{lccc}
		\toprule[0.15em]
		\textbf{} & \textbf{Uncov. Ego} & \textbf{Item/Ego} & \textbf{Friend/Ego} \\ 
		\midrule
		Artists  & 97.4\%	& 1.19  & 1.16 \\
		Movies	& 96.3\% & 1.20 & 1.28\\
		Hashtags & 99.7\%	& 1.78 & 1.79\\
		\bottomrule[0.15em]
		
	\end{tabular}
	\caption{Ego network coverage by dataset. Compared to random, hashtags exhibit the highest locality for all three metrics. Between artists and movies, the metrics are divided.}
	\label{tab:metrics}
\end{table}

Table~\ref{tab:metrics} shows these metrics averaged across all items in each dataset.  All datasets exhibit more locality than random. As with the sparsity and similarity measures, hashtags are more local than movies or artists.  The effect of randomizing by friend versus by item varies by network, indicating that the amount of item-item correlation in the networks is also different.  These results suggest that the dynamics of preferences differ across networks in ways that merit further study. 

\section{Discussion}
We show that on balance, recommendation algorithms that use only ego network information do fairly well compared to algorithms that use an entire dataset, although coverage for ego network-only recommendations may be a concern. 
This result adds evidence to earlier work that argued that recommendations based on only ego networks may be effective in some contexts \cite{sharma11}. For example, when data or computation is local to a mobile device (such as peer to peer recommenders, or mobile recommenders that keep data local to support privacy, as in PocketLens \cite{miller04}), it may be infeasible or undesirable to compute on a large dataset.  Likewise, many websites such as Flixster, TripAdvisor, and CNET 
use existing social networks such as Facebook to support their user accounts.  These sites essentially see individual ego networks drawn from the underlying full Facebook network; our results show that those views may be valuable for recommendation.

We also empirically demonstrate the existence of preference locality using a variety of metrics.
These results align nicely with the stream of work around understanding diffusion in networks, although we focus on preferences rather than memes \cite{leskovec09}. Further, we show that locality differs across preference types and networks.  This suggests the need to examine disparate datasets to support generalizable claims about diffusion processes and their effects on preferences.  

Putting our recommendation and locality results together implicates preference locality as an important reason why using social network information can improve recommendations \cite{jilinchen10,ma11}, as ego-centric recommender performance roughly correlates with preference locality.  Locality isn't the only reason why social information matters---some metrics show higher locality for artists versus movies, and vice versa, despite the higher performance of local \textit{k}-nn for artists over movies---but it does matter.  It seems likely that social network information becomes more valuable for recommendation as locality increases.

Exploring locality and its drivers is a promising direction for future research.
Our proposed metrics are imperfect reflections of locality, capturing some elements that vary across datasets but saying little about why. Further, individual metrics provide different relative orderings of the datasets, suggesting that each captures only part of the story. Properties of the network such as the frequency of item preferences, distribution of friendship links, and visibility of preference information (e.g., through explicit sharing in Twitter or seeing Likes in a Facebook newsfeed) all likely affect the diffusion and locality of preferences. Better models of these properties, along with the endogenous and exogenous processes by which people adopt preferences, would provide both theoretical understanding and practical value for recommender systems.   

\section{Acknowledgements}	
This work was supported by NSF grant 0910664.

\fontsize{9pt}{10pt} \selectfont
\bibliographystyle{aaai}
\bibliography{recco}
\end{document}